\documentclass{article}

\usepackage{arxiv}

\usepackage[utf8]{inputenc} 
\usepackage[T1]{fontenc}    
\usepackage{hyperref}       
\usepackage{url}            
\usepackage{booktabs}       
\usepackage{amsfonts}       
\usepackage{nicefrac}       
\usepackage{microtype}      
\usepackage{lipsum}
\usepackage{graphicx}
\usepackage[warn]{textcomp}

\title{Generalizable prediction of academic performance from short texts on social media}

\author{
  Ivan Smirnov \\
  Institute of Education\\
  National Research University Higher School of Economics\\
  Moscow 101000, Russia \\
  \texttt{ibsmirnov@hse.ru} \\
}

\begin{document}
\maketitle

\begin{abstract}
It has already been established that digital traces can be used to predict various human attributes. In most cases, however, predictive models rely on features that are specific to a particular source of digital trace data. In contrast, short texts written by users -- tweets, posts, or comments -- are ubiquitous across multiple platforms. In this paper, we explore the predictive power of short texts with respect to the academic performance of their authors. We use data from a representative panel of Russian students that includes information about their educational outcomes and activity on a popular networking site, VK. We build a model to predict academic performance from users' posts on VK and then apply it to a different context. In particular, we show that the model could reproduce rankings of schools and universities from the posts of their students on social media. We also find that the same model could predict academic performance from tweets as well as from VK posts. The generalizability of a model trained on a relatively small data set could be explained by the use of continuous word representations trained on a much larger corpus of social media posts. This also allows for greater interpretability of model predictions. 
\end{abstract}


\section*{Introduction}
It is already well-known that various human attributes, from socio-economic status to sexual preferences, can be predicted from digital traces on social media platforms. However, in most cases, predictive models rely on specific features of a particular source of digital trace data. For instance, it was shown in Ref. \cite{kosinski2013private} that a wide range of human traits could be effectively predicted from likes on Facebook, while in Ref. \cite{preoctiuc2015studying} the number of Twitter followers was used (along with other features) to predict users' income. The use of features such as likes or number of followers means that the application of the resulting model is limited to a particular platform, e.g. Facebook or Twitter.

Some forms of digital traces are, in contrast, ubiquitous across multiple platforms. For instance, short texts written by users can be found in the form of tweets, posts on Facebook, comments on message boards, reviews on e-commerce sites, etc. In this paper, we explore the predictive power of short texts with respect to the academic performance of users. First, we build a model that predicts the educational outcomes of users from their posts on a popular Russian social networking site, VK, and then test its predictive performance in different contexts. In particular, we use it to predict the rankings of schools and universities based on the public posts of their students on VK. We also test the generalizability of the model by applying it to users' tweets rather then VK posts.

We use data from a representative Russian panel study titled "Trajectories in Education and Careers" (TrEC) \cite{malik2019russian} that tracks 4,400 students who participated in the Programme for International Student Assessment (PISA) \cite{oecd2014apisa} in 2012. In addition to survey data, this data set contains information about public posts on VK for those participants who agreed to share their VK data (N = 3,483). We use this data to build a model that predicts the PISA scores of users based on the content of their posts. We combine unsupervised learning of word embeddings on a large corpus of VK posts (1.9B tokens) with a simple supervised model trained on individual posts (see Methods for details).

In the next step, we collect data about the academic performance of students from the 100 largest Russian universities and from 914 schools from 3 cities (see Methods). The performance is measured as results of the Unified State Examination (USE), a mandatory examination for all high school graduates in Russia. This performance is available in aggregated form, i.e. for each school the average USE score of its graduates is known and for each university the average USE score of its enrollees is known. We then find VK users who indicated in their profiles that they are from these schools and universities ($N = 154,637$) and apply our model to their public posts. This allows us to rank schools and universities according to their estimated scores from social media posts and then compare the result with an actual ranking based on USE scores.

To further test the generalizability of the model, we select users who indicate a link to their Twitter account in their VK profiles. We download the tweets of these users and check if their academic performance can still be predicted by using the same model applied to tweets instead of VK posts.

Finally, we use our model to explore the variation in the language of users with different educational outcomes. To this end, we compute predicted scores for individual words. This approach is similar to the open-vocabulary method \cite{schwartz2013personality}, however, the use of continuous word representations makes the results less sensitive to the frequency at which a word appears in the training data set. In fact, it even allows computing of reliable scores for words that are absent in the training data set as long as they are frequent enough in the corpus on which word embeddings are trained.

\section*{Methods}
\subsection*{TrEC data}
We used data from the Russian Longitudinal Panel Study of Educational and Occupational Trajectories (TrEC) \cite{malik2019russian}. The study tracks 4,400 students from 42 Russian regions who took the PISA test in 2012 \cite{oecd2014apisa}. We used PISA reading scores as a measure of students' academic performance. PISA defines reading literacy as "understanding, using, reflecting on and engaging with written texts in order to achieve one's goals, to develop one's knowledge and potential, and to participate in society" and considers it a foundation for achievement in other subject areas within the educational system and also a prerequisite for successful participation in most areas of adult life \cite{schleicher2009pisa}. PISA scores are scaled so that the OECD average is 500 and the standard deviation is 100, while every 40 score points roughly correspond to the equivalent of one year of formal schooling \cite{oecd2014apisa}.

In 2018, publicly available information from the social networking site VK was collected for 3,483 TrEC participants who provided informed consent for the use of this data for research purposes. Note that while the initial sample was representative of the 9th-grade high school Russian students in 2012, the social network data is not necessarily representative. There were no publicly available posts for 498 users. The median number of public posts for remaining users was 35. We removed posts that contain URLs from our data set to account for potentially automated postings, and we also excluded re-posts and posts with no text. This resulted in the final data set of 130,575 posts from 2,468 users. 

\subsection*{Model}
We trained the fastText model  \cite{bojanowski2017enriching} on the VK corpus (1.9B tokens, vocabulary size is 2.5M) to obtain vector representations of Russian words (the model is available at \cite{data}). We then represented each post as a 300-dimensional vector by averaging over all its constituent words. We used this representation to train a linear regression model to predict the PISA scores of the posts’ authors. The model was trained on individual posts and the predicted value for users was computed as an average predicted value of all their posts. 

\subsection*{Schools and university data}
VK provides an application programming interface (API) that enables the downloading of information systematically from the site. In particular, downloading user profiles from particular educational institutions and within selected age ranges is possible. For each user, obtaining a list of their public posts is also possible. According to the VK Terms of Service: ``Publishing any content on his/her own personal page, including personal information, the User understands and accepts that this information may be available to other Internet users, taking into account the architecture and functionality of the Site.'' The VK team confirmed that their public API could be used for research purposes. 

We created a list of high schools in Saint Petersburg ($N = 601$), Samara ($N = 214$), and Tomsk ($N = 99$) and then accessed the IDs of users who indicated on VK that they graduated from one of these schools. We removed profiles with no friends from the same school, profiles that already belong to the TrEC data set, and users who indicated several schools in their profiles. The public posts of the remaining users were downloaded and our model was applied to them to get a prediction of the users' academic performance. We then estimated the educational outcomes of a school by averaging the predictions of its students' performance. Overall, 1,064,371 posts from 38,833 users were used at this stage of analysis. The same procedure was performed to obtain a prediction of academic performance for students from the 100 largest universities in Russia ($N_{users} = 115,804$; $N_{posts} = 6,508,770$). Moscow State University was excluded from the analysis as it is known to be a default choice for bots and fake profiles. Even the application of the aforementioned filtering method does not allow reliable data to be obtained for this university, i.e. there is still an order of magnitude more user profiles than the real number of students for a given cohort.

We used data from the web portal "Schools of Saint Petersburg" \cite{spbschools} to obtain the average performance of schools’ graduates in the Unified State Examination (USE). This is a mandatory state examination that all school graduates should pass in Russia. The USE scores for Samara were provided by the web portal "Zeus" \cite{zeus}. The USE scores of the Tomsk schools along with the data on university enrollees \cite{egehse} were collected by the Higher School of Economics.

This information was used to check if the scores predicted from social media data correspond to the ranking of schools and universities based on their USE results. Note that this means that the model was tested not only on a different set of users but that the measure of academic performance was also different from the training settings.

\subsection*{Twitter data}
VK allows users to indicate links to other social media accounts, including Twitter, in their profiles. Only a small proportion of users provide links to their Twitter accounts. In our sample information, about 665 Twitter accounts were available for the Saint-Petersburg data set (less for other cities) and 2,836 Twitter accounts were available for the university data set. This allowed the analysis to be performed only for the university data set. The latest tweets of these 2,836 users were downloaded via Twitter’s API. Note that unlike tweets, VK posts are not limited to 140 or 280 characters. However, most VK posts are short texts (85.2\% of the posts are less than 140 characters and 92.6\% of the posts are less than 280 characters in our sample).

\subsection*{Exploring differential language use}
Thanks to the simplicity of our model, it is possible to obtain meaningful scores for individual words. By construction, the post score is equal to the average scores of its constituent words. This means that it is enough to explore the individual word scores to interpret the model. We ranked all 2.5M words from our dictionary based on the predicted scores and used this ranking to compare the language of users with different educational outcomes. This approach is similar to the open-vocabulary approach \cite{schwartz2013personality,kulkarni2018latent} because it operates on a level of individual words and allows one to find relations that are not captured with traditional closed dictionaries. The main difference is that our approach uses continuous, rather than discrete, word representations and incorporates rich knowledge about the language structure learned from the training of unsupervised word embeddings. 

\section*{Results}
\subsection*{Prediction}
\begin{figure}[h!]
\centering
\includegraphics[width=0.5\linewidth]{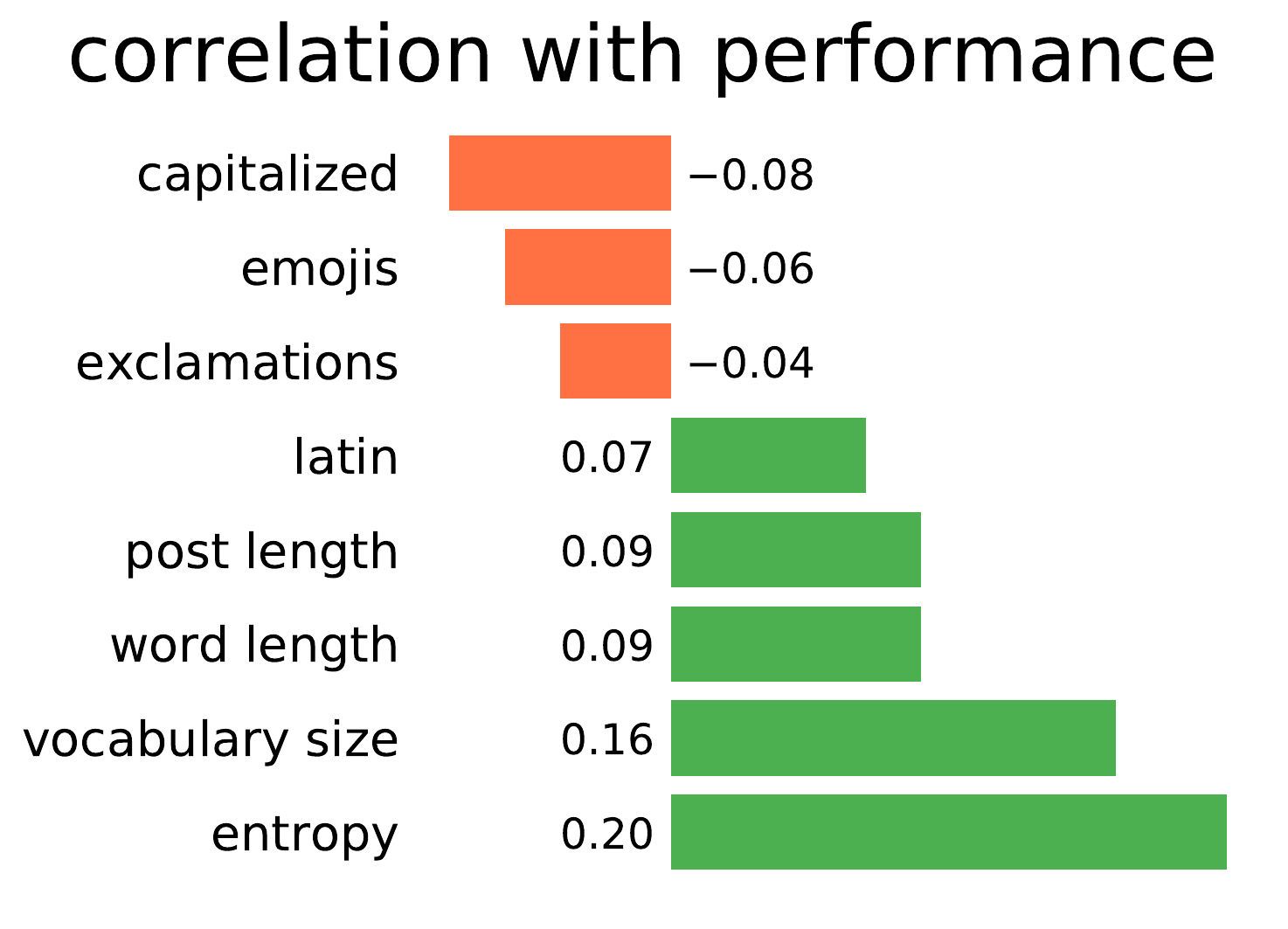}
\caption{\textbf{Pearson correlation between common text features and academic performance.} The use of capitalized words, emojis and exclamations (average number per post normalized by the post length in tokens) is negatively correlated with performance. The use of Latin characters, average post and word length, vocabulary size and entropy of users' texts are positively correlated with academic performance.}
\label{fig:features}
\end{figure}

We first explored the predictive power of common text features with respect to academic performance. We found a small negative effect for the use of capitalized words ($P = 2 \times 10^{-3}$), emojis ($P = 7 \times 10^{-3}$), and exclamations ($P = 0.05$), as seen in Fig. \ref{fig:features}. The use of Latin characters ($P = 5 \times 10^{-3}$), average post length ($P = 2 \times 10^{-4}$), word length ($P = 4 \times 10^{-4}$), and vocabulary size ($P < 10^{-10}$) are positively correlated with academic performance. The strongest correlation was found for the information entropy of users' texts (Pearson's $r = 0.20$, $P < 10^{-15}$).

We used a TF-IDF model to obtain a base-line prediction of academic performance from the users' posts. We selected the 1000 most common unigrams and bigrams from our corpus, excluding stop words, for the Russian language. We then applied a TF-IDF transformation to represent posts as 1000-dimensional vectors and then trained a linear regression model on individual posts to predict the academic performance of their authors. The correlation between predicted and real scores is $r = 0.285$. Here, and for the following models, we report results on the user level obtained using leave-one-out cross-validation, i.e. scores for posts of a certain user were obtained from the model trained on posts of all other users. We obtained significantly better results with a model that used word-embeddings (see Methods). We also find that embeddings trained on the VK corpus outperform models trained on the Wikipedia and Common Crawl corpora (Table~\ref{tab:models}).

\begin{table}[h!]
\centering
\caption{Predictive power of the models measured as Pearson correlation between real and predicted outcomes. The results were computed using leave-one-out cross-validation.}
      \begin{tabular}{lr}
        \hline
        \multicolumn{2}{c}{Correlation coefficient (LOOCV)} \\ \hline
        TF-IDF & $0.284$ \\
        fastText (Wiki) & $0.335$ \\
        fastText (CC) & $0.359$  \\ 
        fastText (VK) & $0.420$  \\ \hline
      \end{tabular}
\label{tab:models}
\end{table}

\begin{figure}[h!]
\centering
\includegraphics[width=0.5\linewidth]{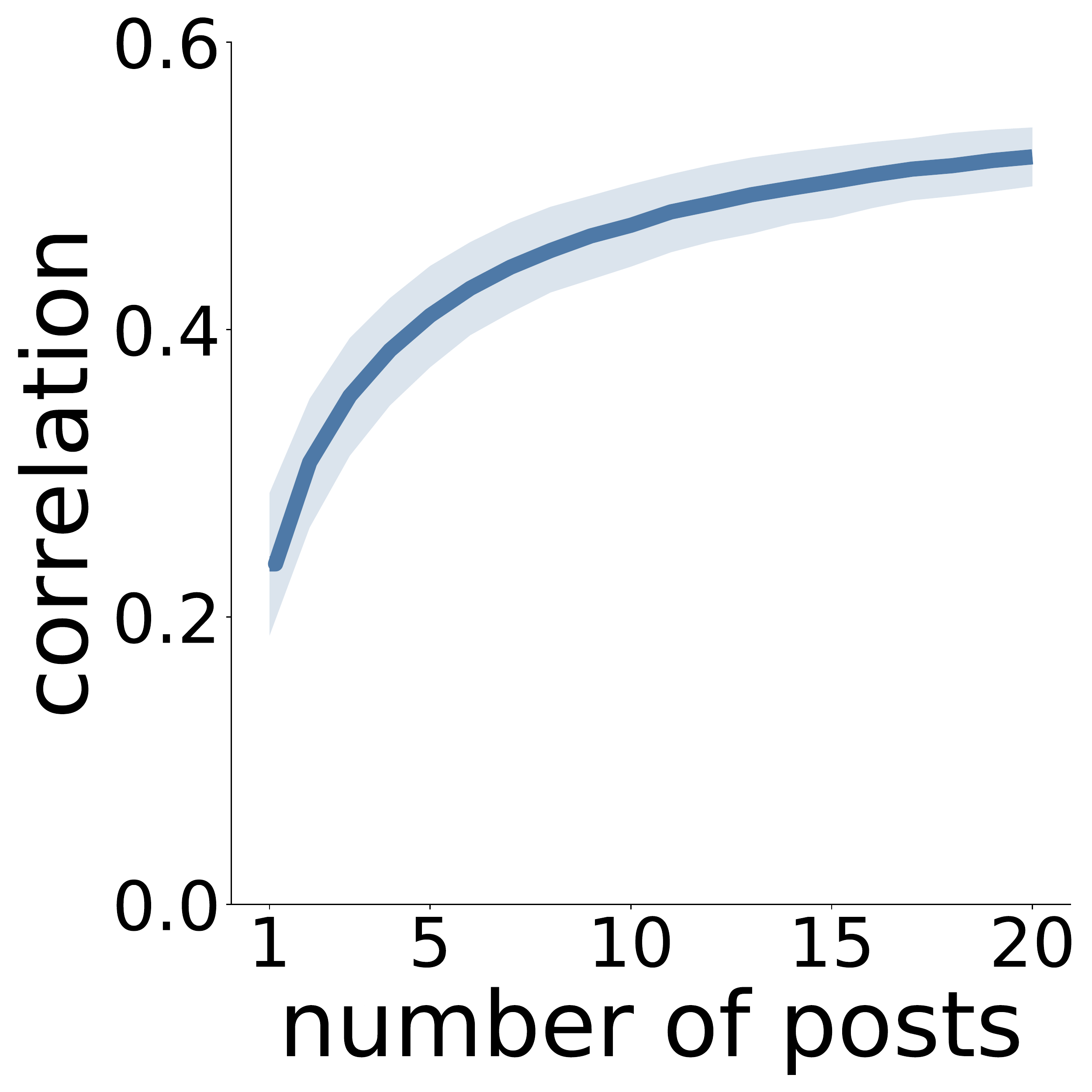}
\caption{\textbf{The predictive power of the model depending on the number of posts used per user.} First, users with at least 20 posts were selected. Then, for each user, $N$ of their posts were selected to predict their academic performance ($N = 1, ..., 20$). The 
shaded region corresponds to the bootstrapped 90\% confidence interval.}
\label{fig:posts}
\end{figure}

The predictive power of a model depends on the number of posts available for each user (see Fig. \ref{fig:posts}). If only one post is available per user, the predictive power is rather low ($r = 0.237$). However, it increases with the number of posts available, reaching $r = 0.541$ for 20 posts per user.

\subsection*{Transfer}
\begin{figure}[h!]
\centering
\includegraphics[width=0.5\linewidth]{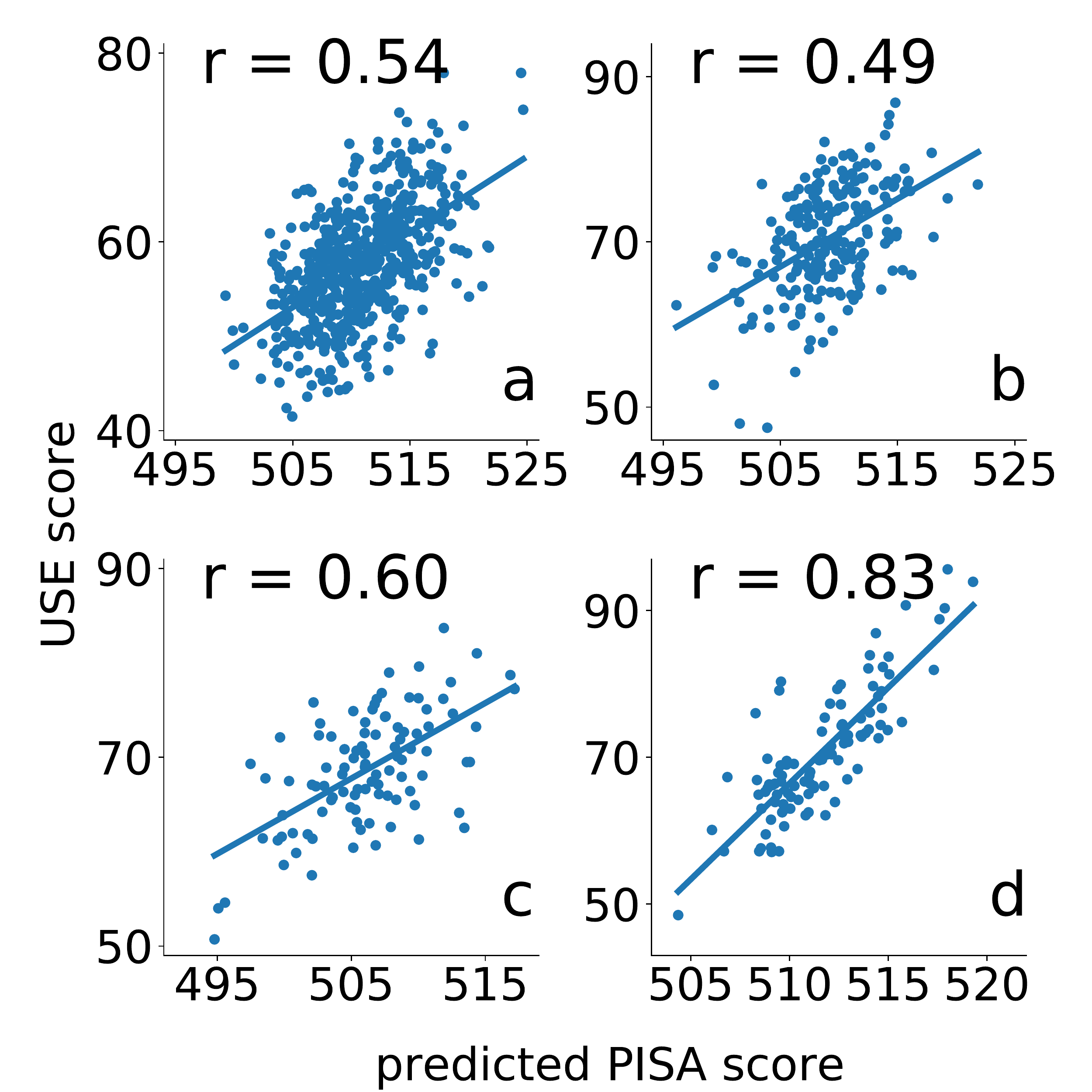}
\caption{\textbf{Correlation between the predicted and real performance of schools and universities.} Pearson's correlation coefficients between predicted school scores and the USE scores of their graduates were computed for Saint-Petersburg (a), Samara (b) and Tomsk (c). The correlation between predicted university scores and the USE scores of their enrollees was also computed for the 100 largest Russian universities (d).}
\label{fig:transfer}
\end{figure}

Figure \ref{fig:transfer} shows the correlation between the predicted performance of schools (a-c) and universities (d) and the USE scores of their graduates or enrollees. In all four cases, we find a relatively strong signal despite the fact that the VK sample might not be representative and that academic performance was measured differently than in training settings, being available only in aggregated form and from secondary sources.

\begin{figure}[h!]
\centering
\includegraphics[width=0.5\linewidth]{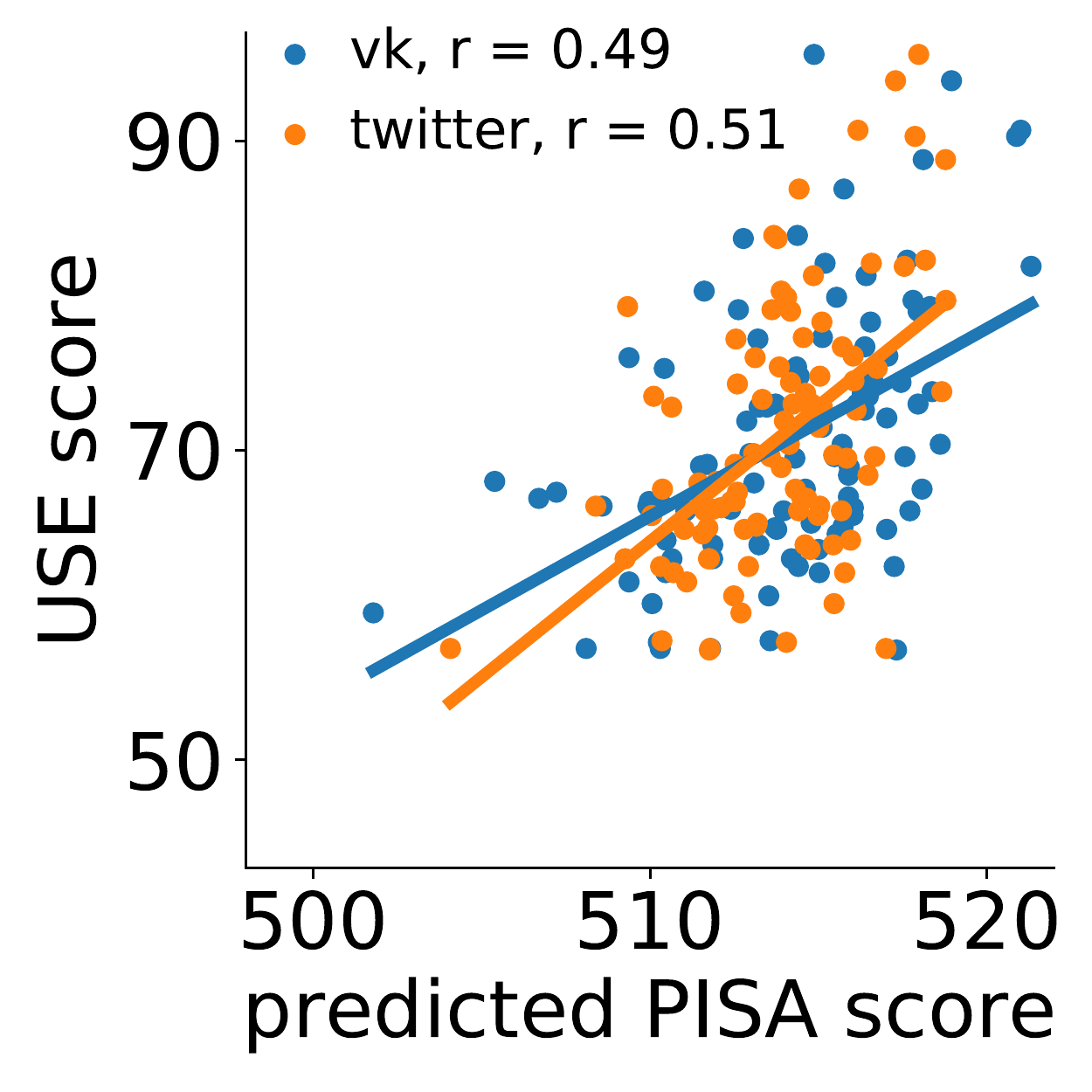}
\caption{\textbf{Comparison of predictions based on VK and Twitter data.} While estimates from Twitter and VK vary for individual universities the overall performance of the model is similar for both cases. Note that the performance of the model is rather low due to the limited number of users per university for whom both VK and Twitter data is available.}
\label{fig:twitter}
\end{figure}

Intriguingly, we find that the substitution of VK posts by tweets doesn't substantially alter the resulting performance (see Fig. \ref{fig:twitter})). For fair comparison, we use VK data only for those users for whom Twitter data was also available. This is why the performance of the model is substantially lower than in Fig. \ref{fig:transfer}d, where all available VK data was used.

\subsection*{Differential language use}
\begin{figure}[h!]
\centering
\includegraphics[width=0.5\linewidth]{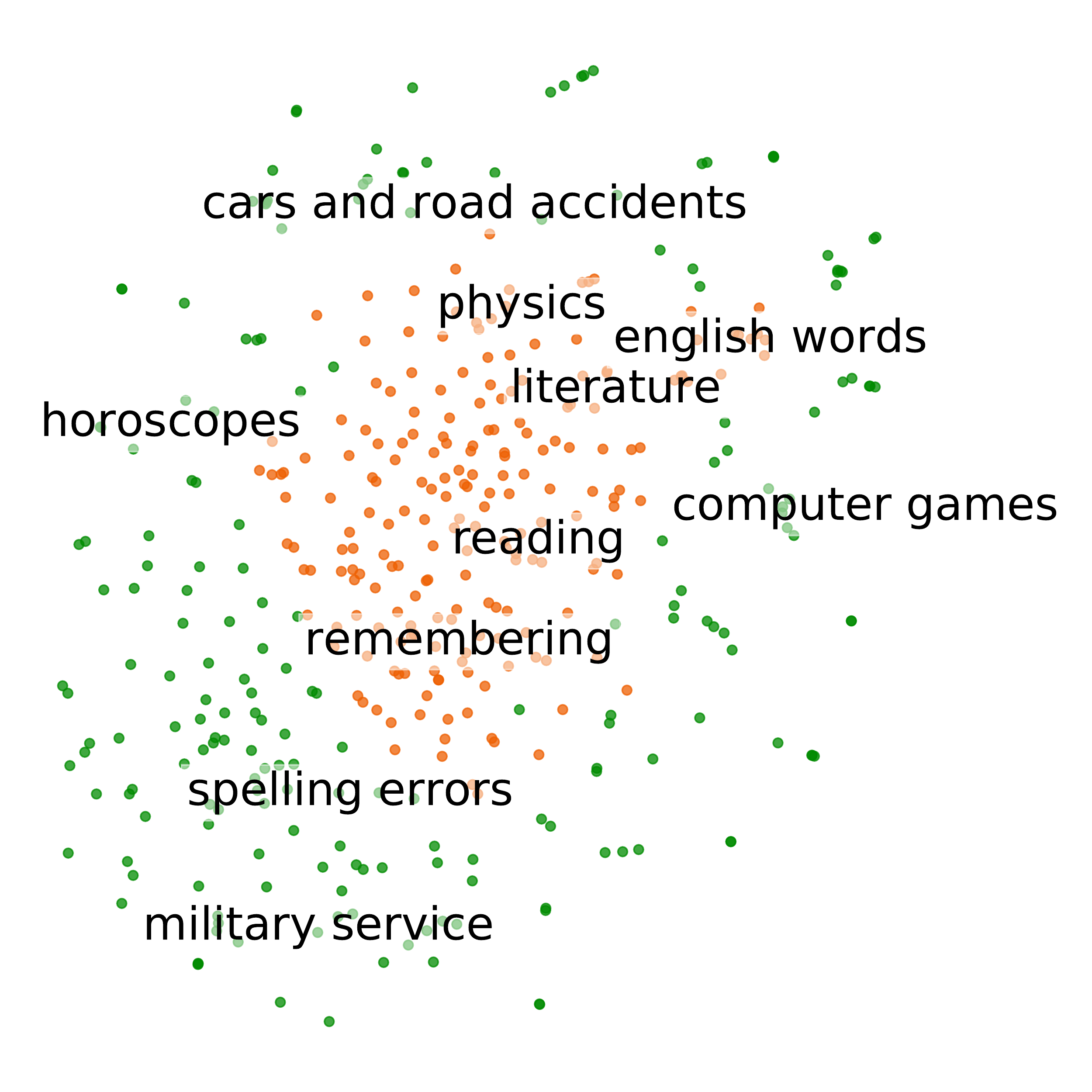}
\caption{\textbf{t-SNE representation of the words with the highest and lowest scores from the training data set.} High performing clusters (orange) include English words and words related to literature, physics, or thinking processes. Low performing clusters (green) include spelling errors and words related to horoscopes, military service, or cars and road accidents.}
\label{fig:words_tsne}
\end{figure}

We explored the resulting model by selecting 400 words with the highest and lowest scores that appear at least 5 times in the training corpus. A t-SNE representation of the embeddings produced by our model helped \cite{maaten2008visualizing} to identify several thematic clusters (Fig. \ref{fig:words_tsne}). High performing clusters include English words (above, saying, yours, must), words related to literature (Dandelion, Bradbury, Fahrenheit, Orwell, Huxley, Faulkner, Nabokov, Brodsky, Camus, Mann, Shelley, Shakespeare), words related to reading (read, reread, published, book, volume), words related to physics (universe, hole, string, theory, quantum, Einstein, Newton, Hawking), and words related to thinking processes including various synonyms of ``thinking'' and ``remembering.''

Low performing clusters include common spelling errors and typos, names of popular computer games, words related to military service (army, to serve, military oath), horoscopes (Aries, Sagittarius), and cars and road accidents (traffic collision, General Administration for Traffic Safety, wheels, tuning).

\begin{figure}[h!]
\centering
\includegraphics[width=0.5\linewidth]{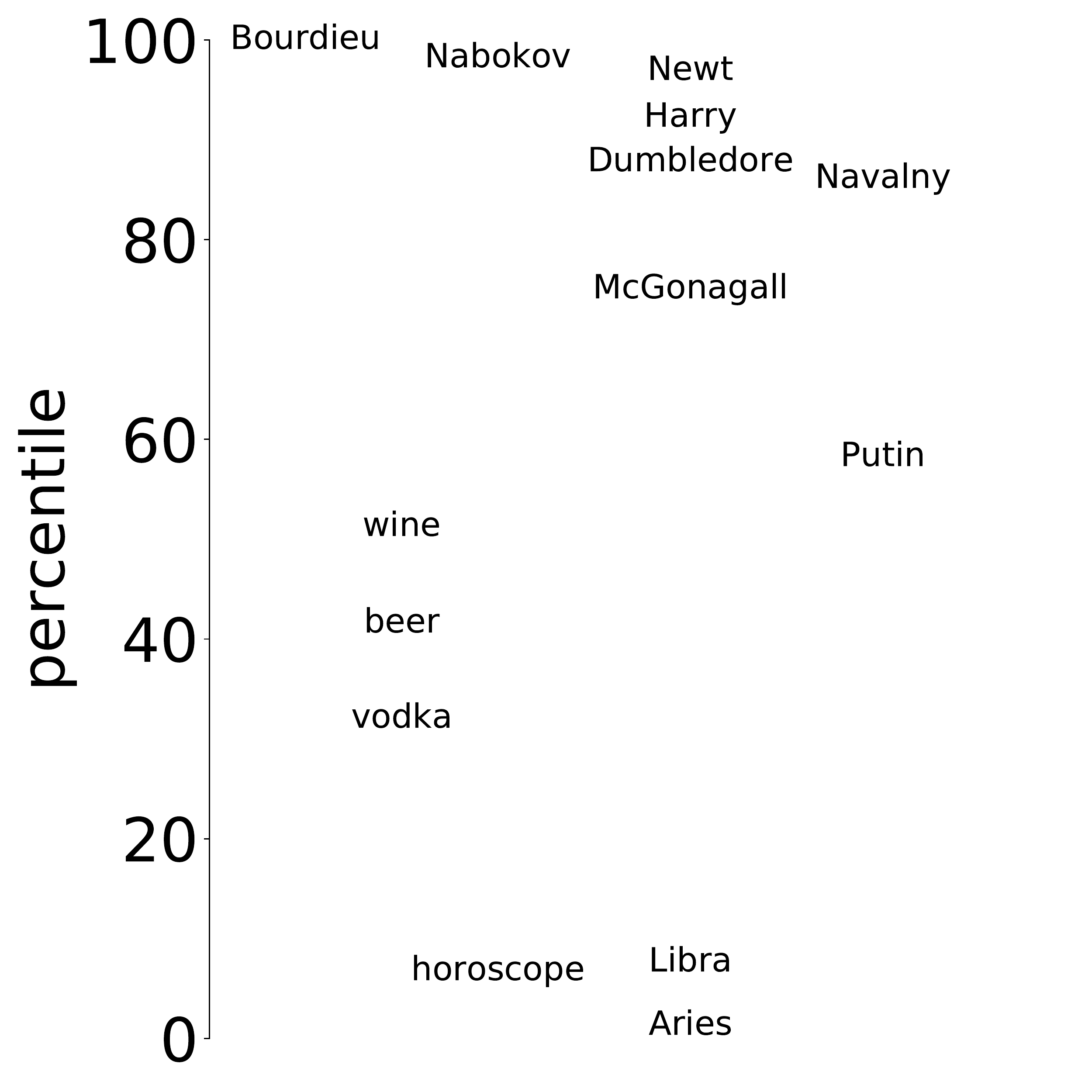}
\caption{\textbf{Ranking of selected words by their predicted score (translated from Russian).} The application of the model to words from different domains confirms its face validity.}
\label{fig:word_percentile}
\end{figure}

The use of continuous word representations allows one to compute scores even for words that were not present in the training data set. We computed the scores for all 2.5M words in our vector model and made it available for further exploration \cite{data}. This could be used for exploratory analysis to get insights into differential use of language with respect to academic performance and could be applied to various domains, from literature to politics or food (Fig. \ref{fig:word_percentile})

\section*{Discussion}
In traditional cross-validation settings, one data set is randomly split into two parts: the model is trained on one part, and then its performance is assessed using the other part. In our case, we use two different data sources (VK and Twitter) and different measures of academic performance (PISA scores and USE scores). Some additional limitations were that the PISA scores were collected in 2012, while most of the posts were written much later, and that the USE scores were available only at the aggregated level. Despite these limitations, we found a relatively strong signal in the data. For instance, we were able to explain 69\% of the variation in universities' scores using information about VK posts of users from these universities. While the result for tweets was significantly lower ($R^2 = 0.26$), this was probably at least partly due to the smaller sample size. Also, note that the prediction of individual scores substantially depends on the number of available posts. If only one post per user is used for prediction, then 6\% of the variation in their academic performance could be explained by our model. This number rises to 29\% if 20 posts are used.

The ability to predict the ranking of educational organizations might seem trivial given that direct ranking information is readily available. However, USE scores are measured only once per cohort, making it extremely hard to estimate any added value provided by an educational organization. A comparison of predicted scores for the same users at different points in time might shed light on factors contributing to the students' progress.

We demonstrate how domain-specific unsupervised learning of word embeddings allows predictive models to be trained using relatively small labeled data sets. One reason is that even words that are rare in the training or testing data sets could be valuable for prediction. For instance, even if the word ``Newt'' never occurs in the training data set, the model could assign a higher score to posts containing it. This would happen if the model learns from the training data set that words from the Harry Potter universe are indicative of high performing students and learns from the unsupervised training that ``Newt'' belongs to this category, i.e. this word is close to other Harry Potter related words in the vector space. This might make the use of continuous word representation preferable to common approaches relying on counting word frequencies. As our approach does not depend on a particular language, source of texts, or target variable (i.e. academic performance could be substituted by income or depression), it could be applied to a wide variety of settings.

Our results also suggest that models trained on text data could be successfully transferred from one data source to another. While this certainly might be useful in some applications, it also means there is a greater risk to users' privacy. If users of platform A do not disclose an attribute X on it, then there is no data to train a model to predict X from digital traces on platform A. However, if X is disclosed on platform B, and both platforms collect short texts from users, then it becomes possible to predict X from digital traces on A given access to data from B. In recent years, face recognition technology has raised particular privacy concerns because of its potential omnipresence and the inability of people to hide from it. In a similar way, digital traces in the form of short texts are ubiquitous, and our results suggest that they allow, if not to identify a person, then at least to predict potentially sensitive private attributes.

\section*{Availability of data and material}
The models and the word rankings are available in the Open Science Framework repository http://doi.org/10.17605/OSF.IO/9PBKR

TrEC data cannot be publicly shared but is available to interested researchers upon request https://trec.hse.ru/en/data

\section*{Competing interests}
The authors declare that they have no competing interests.

\section*{Funding}
This work was supported by a grant from the Russian Science Foundation (project \textnumero 19-18-00271).

\section*{Acknowledgements}
Author thanks the Open Data University Research Consortium \cite{opendata} for providing data on public posts of VK users.

Author thanks Ilyuhin B.V., vice rector for informatization and education quality assessment, TOIPKRO, and the Centre of General and Extracurricular Education, HSE University, for providing data on USE scores of Tomsk schools.

Author thanks Yulia Torgasheva, head of Zeus web-portal for providing data on USE scores of Samara schools.

The TrEC project is supported by the Basic Research Programme of the National Research University Higher School of Economics.

\bibliographystyle{unsrt}  
\bibliography{references} 
\end{document}